\documentclass[english]{article}
\usepackage[T1]{fontenc}
\usepackage[latin9]{inputenc}
\usepackage{graphicx}
\usepackage{esint}

\makeatletter
\newcommand{\lyxaddress}[1]{
\par {\raggedright #1
\vspace{1.4em}
\noindent\par}
}

\makeatother

\usepackage{babel}
\begin{document}

\title{\textbf{Mössbauer rotor experiment as new proof of general relativity:
Rigorous computation of the additional effect of clock synchronization}}

\author{\textbf{Christian Corda}}
\maketitle

\lyxaddress{\textbf{International Institute for Applicable Mathematics and Information
Sciences, B. M. Birla Science Centre, Adarshnagar, Hyderabad 500063
(India); e-mail address: cordac.galilei@gmail.com}}
\begin{abstract}
We received an Honorable Mention at the Gravity Research Foundation
2018 Awards for Essays on Gravitation by showing that a correct general
relativistic interpretation of the Mössbauer rotor experiment represents
a new, strong and independent, proof of Einstein's general theory
relativity (GTR). Here we correct a mistake which was present in our
previous computations on this important issue by deriving a rigorous
computation of the additional effect of clock synchronization. Finally,
we show that some recent criticisms on our general relativistic approach
to the Mössbauer rotor experiment are incorrect, by ultimately confirming
our important result.
\end{abstract}

\section{Introduction}

The Mössbauer effect was discovered by the German physicist R. Mössbauer
in 1958 \cite{key-1}. For this effect, Mössbauer was awarded the
1961 Nobel Prize in Physics. The effect consists in resonant and recoil-free
emission and absorption of gamma rays, without loss of energy, by
atomic nuclei bound in a solid. It is very important for various research
fields in physics and chemistry. In this paper, the so called Mössbauer
rotor experiment will be discussed, see Figure 1. In that case, the
Mössbauer effect works by using an absorber orbited around a source
of resonant radiation (or vice versa). The aim of the experiment is
to verify the relativistic time dilation for a moving resonant absorber
which generates a relative energy shift between emission and absorption
lines. 
\begin{figure}
\includegraphics{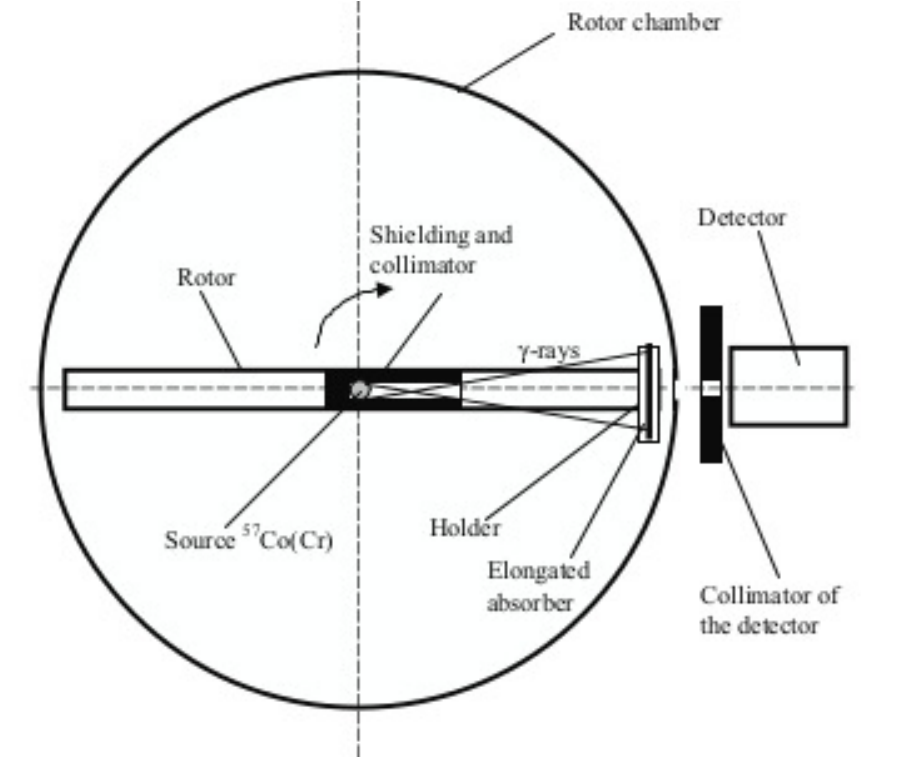}

\caption{Scheme of the new Mössbauer rotor experiment, adapted from ref. \cite{key-2}}
\end{figure}

The idea to use a general relativistic framework in order to explain
the Mössbauer rotor experiment has a long history, which started from
the famous book of Pauli \cite{key-3}. The key point is \emph{Einstein's
equivalence principle}, (EEP) which states the equivalence between
the gravitational \textquotedbl{}force\textquotedbl{} and the \emph{pseudo-force}
experienced by an observer in a non-inertial frame of reference {[}4
- 6{]}. The case of the rotating frame of reference of the Mössbauer
rotor experiment is a particular case of the EEP {[}4 - 6{]}. This
permits one to reanalyse the theoretical framework of the Mössbauer
rotor experiment directly in the rotating frame of reference through
a full general relativistic treatment {[}4 - 6{]}. 

A historical experiment on the Mössbauer rotor effect was due to Kündig
\cite{key-7}. Kündig's work was recently reanalysed by a team of
researchers in \cite{key-2,key-8}. Such a team, first reanalysed
in \cite{key-8} the data of Kündig's experiment. Then, the team's
researchers realized their own experiment on the time dilation effect
in a rotating system \cite{key-2}. In \cite{key-8}, the authors
found that in the original experiment of Kündig \cite{key-7} errors
were present in the data processing. After the correction of the errors
of Kündig, the experimental data gave the value \cite{key-8}

\begin{equation}
\frac{\nabla E}{E}\simeq-k\frac{v^{2}}{c^{2}},\label{eq: k}
\end{equation}
where $k=0.596\pm0.006$, instead of the standard general relativistic
prediction $k=0.5$ due to time dilation \cite{key-3}. This was a
puzzling issue. In \cite{key-8}, the authors emphasized that, on
one hand, the deviation of the coefficient $k$ in equation (\ref{eq: k})
from $0.5$ exceeds by almost 20 times the measuring error. On the
other hand, the revealed deviation cannot be attributed to the influence
of rotor vibrations and/or other kinds of disturbing factors. The
potential disturbing factors was indeed excluded by a very good methodology
of Kündig \cite{key-7}. Such a methodology is given by a first-order
Doppler modulation of the energy of $\gamma-$quanta on a rotor at
each fixed rotation frequency \cite{key-7}. Hence, the experiment
of Kündig is today considered as being the most precise among other
similar experiments {[}9 - 13{]}. The experimenters {[}9 - 13{]} indeed
measured only the count rate of detected $\gamma-$quanta as a function
of rotation frequency. In \cite{key-8} it has also been shown that
the experiment in \cite{key-13} confirms the supposition $k>0.5.$
Remarkably, the experiment in \cite{key-13} contains much more data
than the ones in {[}9 - 12{]}. In order to better investigating the
results in \cite{key-8}, the authors realized their own experiment
\cite{key-2}. In \cite{key-2}, neither the scheme of the Kündig
experiment \cite{key-7}, nor the schemes of other known experiments
on the subject previously mentioned above {[}9 - 13{]} have been repeated.
This permitted to obtain a completely independent information on the
value of $k$ in Eq. (\ref{eq: k}). The authors refrained indeed
from the first-order Doppler modulation of the energy of $\gamma-$quanta
\cite{key-2}. This permitted to exclude the uncertainties in the
realization of this method \cite{key-2}. The standard scheme in {[}9
- 13{]} has been followed also in \cite{key-2}. It means that the
count rate of detected $\gamma-$quanta $N$ as a function of the
rotation frequency $\nu$ has been measured. But, differently from
the experiments {[}9 - 13{]}, in \cite{key-2} the influence of chaotic
vibrations on the measured value of $k$ has been evaluated. A method
involving a joint processing of the data collected for two selected
resonant absorbers with the specified difference of resonant line
positions in the Mössbauer spectra has been developed \cite{key-2}.
The final result was the value $k=0.68\pm0.03$ \cite{key-2}. This
confirms that the coefficient $k$ in Eq. (\ref{eq: k}) substantially
exceeds 0.5. The reader can see the scheme of the new Mössbauer rotor
experiment in Figure 1. Further technical details on this scheme can
be found in \cite{key-2}. 

In {[}4 - 6{]} we reanalysed the theoretical framework of the Mössbauer
rotor experiment directly in the rotating frame of reference through
a full general relativistic treatment. We have shown that, in previous
analyses in the literature, an important effect of clock synchronization
has been missed. Thus, the correct general relativistic prevision
gives $k\simeq\frac{2}{3}$ {[}4 - 6{]}, which is in perfect agreement
with the new experimental results in \cite{key-2}. In other words,
the general relativistic interpretation {[}4 - 6{]} shows that the
new experimental results of the Mössbauer rotor experiment in \cite{key-2}
are a new, strong and independent, proof of the GTR. Remarkably, our
results on the Mössbauer rotor experiment received an Honorable Mention
at the Gravity Research Foundation 2018 Awards for Essays on Gravitation
\cite{key-4}. 

We must also stress that various papers in the literature (included
ref. \cite{key-9} published in Phys. Rev. Lett.) missed the additional
effect of clock synchronization {[}2, 7 - 16{]}. This generated some
claim of invalidity of relativity theory and/or some attempts to explain
the experimental results through non-conventional or ``exotic''
effects \cite{key-2,key-8,key-14,key-15,key-16}. One of these ``exotic''
effects will be partially discussed in Section 3 of this paper.

Here, an important mistake which was present in our previous computations
on this important issue in {[}4 - 6{]} will be solved, by deriving
a rigorous computation of the additional effect of clock synchronization.
Finally, it will be shown that some recent criticisms \cite{key-17,key-18}
on our general relativistic approach to the Mössbauer rotor experiment
are incorrect, by ultimately confirming our important result.

\section{General relativistic interpretation of the Mössbauer rotor experiment }

\subsection{The ``gravitational blueshift''}

Following {[}4 - 6{]}, one uses a transformation from an inertial
frame, in which the space-time is Minkowskian, to a rotating frame
of reference in cylindrical coordinates. In the starting inertial
frame, one has the line-element {[}4 - 6{]}

\begin{equation}
ds^{2}=c^{2}dt^{2}-dr^{2}-r^{2}d\phi^{2}-dz^{2}.\label{eq: Minkowskian}
\end{equation}
One considers the following transformation to a frame of reference
$\left\{ t',r',\phi'z'\right\} $which rotates with an uniform angular
rate $\omega$ with respect to the starting inertial frame {[}4 -
6{]}

\begin{equation}
\begin{array}{cccc}
t=t'\; & r=r' & \;\phi=\phi'+\omega t'\quad & z=z'\end{array}.\label{eq: trasformazione Langevin}
\end{equation}
Eq. (\ref{eq: Minkowskian}) becomes the famous Langevin metric in
the rotating frame {[}4 - 6{]}
\begin{equation}
ds^{2}=\left(1-\frac{r'^{2}\omega^{2}}{c^{2}}\right)c^{2}dt'^{2}-2\omega r'^{2}d\phi'dt'-dr'^{2}-r'^{2}d\phi'^{2}-dz'^{2}.\label{eq: Langevin metric}
\end{equation}
Despite it is simple to grasp, the transformation (\ref{eq: trasformazione Langevin})
is highly illustrative of the GTR general covariance. It indeed shows
that one can start to work in a \textquotedbl{}simpler\textquotedbl{}
frame and then transforming to a more \textquotedbl{}complex\textquotedbl{}
one {[}4 - 6{]}. Through the EEP, the line element (\ref{eq: Langevin metric})
is interpreted in terms of a curved space-time in presence of a static
gravitational field {[}4 - 6{]}. This gives a pure general relativistic
interpretation of the pseudo-force that an observer in a rotating,
non-inertial frame of reference experiences {[}4 - 6{]}. One sets
the origin of the reference frames (both of the rotating and resting
ones) in the source of the emitting radiation {[}4 - 6{]}. Thus, one
gets a first contribution arising from the ``gravitational blueshift''
{[}4 - 6{]}. One easily computes this contribution by using Eq. (25.26)
in \cite{key-19}, which, in the twentieth printing 1997 of \cite{key-19},
reads
\begin{equation}
z\equiv\frac{\Delta\lambda}{\lambda}=\frac{\lambda_{received}-\lambda_{emitted}}{\lambda_{emitted}}=|g_{00}(r'_{1})|^{-\frac{1}{2}}-1.\label{eq: z  MTW}
\end{equation}
This equation gives the redshift of a photon emitted by an atom at
rest in a gravitational field and received by an observer at rest
at infinity. Here, a slightly different equation with respect to Eq.
(25.26) in \cite{key-19} will be used. In fact, here one considers
a gravitational field which increases with increasing radial coordinate
$r'$. Instead, Eq. (25.26) in \cite{key-19} concerns a gravitational
field which decreases with increasing radial coordinate {[}4 - 6{]}.
The result will be a blueshift (a negative shift) rather than a redshift.
In addition, we set the zero potential in $r'=0$ rather than at infinity,
and we use the proper time $\tau$ rather than the wavelength $\lambda$
{[}4 - 6{]}. Therefore, from Eq. (\ref{eq: Langevin metric}), one
gets {[}4 - 6{]}
\begin{equation}
\begin{array}{c}
z_{1}\equiv\frac{\Delta\tau_{10}-\Delta\tau_{11}}{\tau_{1}}=1-|g_{00}(r'_{1})|{}^{-\frac{1}{2}}=1-\frac{1}{\sqrt{1-\frac{\left(r'_{1}\right)^{2}\omega^{2}}{c^{2}}}}\\
\\
=1-\frac{1}{\sqrt{1-\frac{v^{2}}{c^{2}}}}\simeq-\frac{1}{2}\frac{v^{2}}{c^{2}}.
\end{array}\label{eq: gravitational redshift}
\end{equation}
Here, $\Delta\tau_{10}$ is the delay of the emitted radiation, $\Delta\tau_{11}$
is the delay of the received radiation, $r'_{1}\simeq c\tau_{1}$
is the radial coordinate of the detector and $v=r'_{1}\omega$ is
the tangential velocity of the detector {[}4 - 6{]}. For the sake
of completeness, we clarify the meaning of ``delay of the emitted
and of the received radiation''. We are referring to the issue that
the proper time interval between events of emission of two photons
as measured by the standard clock at the point of emission is different
from the proper time interval between events of absorption of those
photons as measured by identical standard clock at the point of absorption.
This was the initial reasoning of Einstein in \cite{key-28} and was
elegantly adopted by Weyl in \cite{key-29}.

Then one obtains the first contribution, say $k_{1}=\frac{1}{2}$,
to $k$ {[}4 - 6{]}. Let us again emphasize that it is the power of
the EEP which enabled us to use a pure general relativistic treatment
in previous analysis {[}4 - 6{]}.

\subsection{Additional effect of clock synchronization: correcting an important
mistake}

In order to understand the necessity of an additional effect, one
stresses that the variations of proper time $\Delta\tau_{10}$ and
$\Delta\tau_{11}$ have been calculated in the origin of the rotating
frame which is located in the source of the radiation {[}4 - 6{]}.
Then, the key point is that the detector moves with respect to the
origin in the rotating frame {[}4 - 6{]}. This means that the clock
in the detector, which is located in the laboratory frame at rest,
has to be synchronized with the rotating clock in the origin. This
generates a second, additional, contribution to time dilation {[}4
- 6{]}, which was missed in previous analyses {[}2, 3, 7 - 18{]}.
In order to compute this second contribution, in {[}4 - 6{]} we claimed
to use Eq. (10) of \cite{key-20}, by stressing that it represents
the proper time increment $d\tau$ on the moving clock having radial
coordinate $r'$ for values $v\ll c$. Actually, in our works {[}4
- 6{]} an important mistake was present. In fact, Eq. (10) of \cite{key-20}
is given by 

\begin{equation}
d\tau=dt'-\frac{\omega r'^{2}d\phi'}{c^{2}}\label{eq:secondo contributo}
\end{equation}
while in {[}4 - 6{]} we used 
\begin{equation}
d\tau=dt'\left(1-\frac{r'^{2}\omega^{2}}{c^{2}}\right).\label{eq:secondo contributo sbagliata}
\end{equation}
Now, Eqs. (\ref{eq:secondo contributo}) and (\ref{eq:secondo contributo sbagliata})
coincide each other only if $\omega=\frac{d\phi'}{dt'}.$ But this
cannot happen because we consider light propagating in the radial
direction, which implies $d\phi'=dz'=0$. The correct way to proceed
is the following. One recalls that $t'$ represents the time coordinate
in the rotating frame, but, from Eq. (\ref{eq: trasformazione Langevin})
it is also $t'=t$, where $t$ represents both of the coordinate time
and the proper time of the Minkowskian laboratory frame. In a gravitational
field, the rate $d\tau$ of the proper time is related to the rate
$dt'$ of the coordinate time by \cite{key-21}
\begin{equation}
d\tau^{2}=g_{00}dt'^{2}.\label{eq: relazione temporale}
\end{equation}
From the first of Eqs. (\ref{eq: trasformazione Langevin}), i.e.
$t=t',$ one gets immediately from Eq. (\ref{eq: relazione temporale})
\begin{equation}
d\tau^{2}=g_{00}dt{}^{2}.\label{eq: relazione tempi propri}
\end{equation}
Thus, as in a Minkowskian space-time the proper time is equal to the
coordinate time, Eq. (\ref{eq: relazione tempi propri}) gives the
difference of proper time between the laboratory frame and the rotating
frame. 

From Eq. (\ref{eq: Langevin metric}) it is $g_{00}=\left(1-\frac{r'^{2}\omega^{2}}{c^{2}}\right),$
and one can, in turn, rewrite Eq. (\ref{eq: relazione temporale})
as 
\begin{equation}
c^{2}d\tau^{2}=\left(1-\frac{r'^{2}\omega^{2}}{c^{2}}\right)c^{2}dt'^{2}.\label{eq: relazione temporale 2}
\end{equation}
Then, using again Eq. (\ref{eq: trasformazione Langevin}), one gets
\begin{equation}
c^{2}dt'^{2}=c^{2}dt{}^{2}=dr^{2}=dr'^{2},\label{eq: obviously}
\end{equation}
where the equality 
\begin{equation}
c^{2}dt{}^{2}=dr^{2}\label{eq: equality}
\end{equation}
depends on the issue that light propagates in the radial direction
in the laboratory frame (the source is indeed a rest in that frame).
Thus, one has $d\phi=dz=0$ in Eq. (\ref{eq: Minkowskian}) and, by
inserting the condition of null geodesics $ds=0$ in the same equation,
one immediately obtains Eq. (\ref{eq: equality}). 

Hence, Eq. (\ref{eq: relazione temporale 2}) becomes 
\begin{equation}
c^{2}d\tau^{2}=\left(1-\frac{r'^{2}\omega^{2}}{c^{2}}\right)dr'^{2}.\label{eq: relazione temporale 3}
\end{equation}
For the sake of correctness, we stress that we considered only radial
propagation of light in the rotating system in our previous works
{[}4 - 6{]}. Actually, this is not a mistake. In fact, despite propagation
of light in the rotating frame is not radial, the ``gravitational
field'' has pure radial direction. This implies that the momentum
of photons in the rotation direction, which is perpendicular to the
radial direction, is conserved. For the same reason, also the momentum
of photons in the z-direction is conserved. As a consequence, the
two blueshift effects works only in the radial direction in the rotating
frame. In fact, despite propagation of light in the rotating frame
is not radial, the formula which governs the effect of clock synchronization,
i.e. Eq. (\ref{eq: relazione temporale 3}), depends only on the radial
coordinate in the rotating frame. 

The root square of Eq. (\ref{eq: relazione temporale 3}) is 
\begin{equation}
cd\tau=\sqrt{1-\frac{r'^{2}\omega^{2}}{c^{2}}}dr',\label{eq: secondo contributo finale}
\end{equation}
which is equal to Eq. (10) in \cite{key-4}. Thus, now one can follow
step by step the analysis in {[}4 - 6{]}. 

Eq. (\ref{eq: secondo contributo finale}) is well approximated by
{[}4 - 6{]}
\begin{equation}
cd\tau\simeq\left(1-\frac{1}{2}\frac{r'^{2}\omega^{2}}{c^{2}}+....\right)dr',\label{eq: well approximated}
\end{equation}
which gives the second contribution of order $\frac{v^{2}}{c^{2}}$
to the variation of proper time {[}4 - 6{]}
\begin{equation}
c\Delta\tau_{2}=\int_{0}^{r'_{1}}\left(1-\frac{1}{2}\frac{\left(r'_{1}\right)^{2}\omega^{2}}{c^{2}}\right)dr'-r'_{1}=-\frac{1}{6}\frac{\left(r'_{1}\right)^{3}\omega^{2}}{c^{2}}=-\frac{1}{6}r'_{1}\frac{v^{2}}{c^{2}}.\label{eq: delta tau 2}
\end{equation}
One recalls that $r'_{1}\simeq c\tau$ is the radial distance between
the source and the detector. Then, one gets the second contribution
of order $\frac{v^{2}}{c^{2}}$ to the blueshift as {[}4 - 6{]}
\begin{equation}
z_{2}\equiv\frac{\Delta\tau_{2}}{\tau_{1}}=-k_{2}\frac{v}{c^{2}}^{2}=-\frac{1}{6}\frac{v^{2}}{c^{2}}.\label{eq: z2}
\end{equation}
Then, one obtains $k_{2}=\frac{1}{6}$ and, using Eqs. (\ref{eq: gravitational redshift})
and (\ref{eq: z2}), the total blueshift is {[}4 - 6{]}
\begin{equation}
\begin{array}{c}
z\equiv z_{1}+z_{2}=\frac{\Delta\tau_{10}-\Delta\tau_{11}+\Delta\tau_{2}}{\tau_{1}}=-\left(k_{1}+k_{2}\right)\frac{v^{2}}{c^{2}}\\
\\
=-\left(\frac{1}{2}+\frac{1}{6}\right)\frac{v^{2}}{c^{2}}=-k\frac{v^{2}}{c^{2}}=-\frac{2}{3}\frac{v^{2}}{c^{2}}=-0.\bar{6}\frac{v^{2}}{c^{2}},
\end{array}\label{eq: z totale}
\end{equation}
which is completely consistent with the experimental result $k=0.68\pm0.03$
in \cite{key-2}. In Eq. (\ref{eq: z totale}) $\nabla\tau_{10}$
is the delay of the emitted radiation, $\Delta\tau_{11}$ is the delay
of the received radiation and $\Delta\tau_{2}$ is the delay due to
clock synchronization. 

We can check our computation as it follows. Inserting the condition
of null geodesics $ds=0$ in Eq. (\ref{eq: Langevin metric}) and
considering pure radial motion of photons (as we stressed above, considering
only radial propagation of light in the rotating system is not a mistake
because of the conservation of the momentum of photons in both the
direction of rotation and the z-direction) one gets {[}4 - 6{]}
\begin{equation}
cdt'=\frac{dr'}{\sqrt{1-\frac{r'^{2}\omega^{2}}{c^{2}}}},\label{eq: tempo 2}
\end{equation}
where the positive sign in the square root has been taken, because
the radiation is propagating in the positive $r$ direction {[}4 -
6{]}. Eq. (\ref{eq: tempo 2}) represents the variation of the coordinate
time with respect to the radial coordinate in the rotating frame.
But we again recall that, from Eq. (\ref{eq: trasformazione Langevin}),
one gets also $t'=t$, being $t$ both of the coordinate time and
the proper time $\tau$ of the laboratory frame. Eq. (\ref{eq: tempo 2})
is well approximated by 
\begin{equation}
cd\tau\simeq\left(1+\frac{1}{2}\frac{r'^{2}\omega^{2}}{c^{2}}+....\right)dr',\label{eq: well approximated 2}
\end{equation}
which permits to obtain 
\begin{equation}
c\Delta\tau_{3}=\int_{0}^{r'_{1}}\left(1+\frac{1}{2}\frac{\left(r'_{1}\right)^{2}\omega^{2}}{c^{2}}\right)dr'-r'_{1}=+\frac{1}{6}\frac{\left(r'_{1}\right)^{3}\omega^{2}}{c^{2}}=+\frac{1}{6}r'_{1}\frac{v^{2}}{c^{2}}.\label{eq: delta tau 3}
\end{equation}
Thus, Eq. (\ref{eq: delta tau 3}) shows that, if the rotating observer
measures an additional (negative) proper time $\Delta\tau_{2}$ given
by Eq. (\ref{eq: delta tau 2}), the observer in the laboratory frame
measures an additional proper time $\Delta\tau_{3}$ given by Eq.
(\ref{eq: delta tau 3}) which is 
\begin{equation}
\Delta\tau_{3}=-\Delta\tau_{2},\label{eq: tempi propri}
\end{equation}
as one expects.

\subsection{Correspondence with the use of the GTR in Global Positioning Systems }

For the sake of completeness, in this Subsection we review a discussion
in \cite{key-5} on the correspondence between the analysis of the
Mössbauer rotor experiment and the use of the GTR in Global Positioning
Systems (GPS).

The additional factor $-\frac{1}{6}$ in eq. (\ref{eq: z2}) arises
from clock synchronization \cite{key-5}. This means that its theoretical
absence in {[}2, 3, 7 - 18{]} is due to the incorrect comparison of
clock rates between a clock at the origin and one at the detector
\cite{key-5}. This generated wrong claims of invalidity of relativity
theory and/or some attempts to explain the experimental results through
``exotic'' effects \cite{key-2,key-8,key-14,key-15,key-16} which,
instead, must be rejected. One of these ``exotic'' effects will
be partially discussed in next Section of this paper.

We used \cite{key-20} for introducing a discussion of the Langevin
metric. This also concerns the use of the GTR in GPS and permits one
to realize the following \cite{key-5}. The additional term $-\frac{1}{6}$
in Eq. (\ref{eq: z2}) is similar to the correction that one has to
consider in GPS when one accountes for the difference between the
time measured in a frame co-rotating with the Earth geoid and the
time measured in a non-rotating Earth centered frame, which is locally
inertial \cite{key-5} (and also the difference between the proper
time of an observer at the surface of the Earth and at infinity).
In fact, by simply considering the redshift due to the Earth gravitational
field, but neglecting the effect of the Earth's rotation, GPS cannot
work \cite{key-5}. Following \cite{key-5,key-20}, if one wants to
address the problem of clock synchronization within the GPS, one starts
from an approximate solution of Einstein\textquoteright s field equations
in isotropic coordinates in a locally inertial, non-rotating, freely
falling coordinate system with origin at the Earth's center of mass
\cite{key-5,key-20}
\begin{equation}
ds^{2}=\left(1+\frac{2V}{c^{2}}\right)\left(cdt\right)^{2}-\left(1-\frac{2V}{c^{2}}\right)\left(dr^{2}+r^{2}d\theta^{2}+r^{2}\sin^{2}\theta d\phi^{2}\right)\label{eq: approximate solution}
\end{equation}
where $V$ is the Newtonian gravitational potential of the Earth and
${r,\theta,\phi}$ are spherical polar coordinates. $V$ is approximately
\cite{key-5,key-20}
\begin{equation}
V\simeq\frac{-GM_{E}}{r}\left[1-J_{2}\left(\frac{a_{1}}{r}\right)^{2}P_{2}\cos\theta\right],\label{eq: V}
\end{equation}
where $M_{E}$ is the Earth's mass, $J_{2}$ the Earth's quadrupole
moment coefficient, $a_{1}$ the Earth's equatorial radius and $P_{2}$
the Legendre polynomial of degree $2$ \cite{key-5,key-20}. One retains
only terms of first order in the small quantity $\frac{V}{c^{2}}$
in Eq. (\ref{eq: approximate solution}). In fact, higher multipole
moment contributions to Eq. (\ref{eq: V}) have negligible effect
for relativity in GPS \cite{key-5,key-20}. The equivalent transformations
of Eqs. (\ref{eq: trasformazione Langevin}) for spherical polar coordinates
read \cite{key-5,key-20} 
\begin{equation}
t=t',\;r=r',\;\theta=\theta',\;\phi'+\omega_{E}t'\label{eq: Langevin sferiche}
\end{equation}
where $\omega_{E}$ is the Earth's uniform angular rate. If one applies
the transformations (\ref{eq: Langevin sferiche}) to the line element
(\ref{eq: approximate solution}) and retains only terms of order
$1/c^{2}$, the line element for the so called Earth-centered, Earth-fixed,
reference frame (ECEF frame) is obtained as \cite{key-5,key-20} 
\begin{equation}
\begin{array}{c}
ds^{2}=\left[1+\frac{2V}{c^{2}}-\left(\frac{\omega_{E}r'\sin\theta'}{c}\right)^{2}\right]\left(cdt'\right)^{2}+2\omega_{E}r'^{2}\sin^{2}\theta'd\phi'dt'\\
\\
-\left(1-\frac{2V}{c^{2}}\right)\left(dr'^{2}+r'^{2}d\theta'^{2}+r'^{2}\sin^{2}\theta'd\phi'^{2}\right).
\end{array}\label{eq: ECEF frame}
\end{equation}
Standard clocks at rest at infinity define the rate of coordinate
time in Eq. (\ref{eq: approximate solution} \cite{key-5,key-20}.
Instead, one prefers considering the rate of coordinate time by standard
clocks at rest on the Earth's surface \cite{key-5,key-20}. Thus,
one introduces a new coordinate time $t''$ through a constant rate
change \cite{key-5,key-20}:
\begin{equation}
t''=\left(1+\frac{\Phi_{0}}{c^{2}}\right)t'=\left(1+\frac{\Phi_{0}}{c^{2}}\right)t,\label{eq: constant rate change}
\end{equation}
where \cite{key-5,key-20}
\begin{equation}
\Phi_{0}\equiv-\frac{GM_{E}}{a_{1}}-\frac{GM_{E}J_{2}}{2a_{1}}-\frac{1}{2}\left(\omega_{E}a_{1}\right)^{2}.\label{eq: fi zero}
\end{equation}
The correction (\ref{eq: constant rate change}) is order seven parts
in $10^{10}$ \cite{key-5,key-20}. If one applies this time scale
change in the ECEF line element (\ref{eq: ECEF frame}) and retains
only terms of order $\frac{1}{c^{2}}$, one obtains \cite{key-5,key-20}
\begin{equation}
\begin{array}{c}
ds^{2}=\left[1+\frac{2\left(\Phi-\Phi_{0}\right)}{c^{2}}\right]\left(cdt''\right)^{2}+2\omega_{E}r'^{2}\sin^{2}\theta'd\phi'dt''\\
\\
-\left(1-\frac{2V}{c^{2}}\right)\left(dr'^{2}+r'^{2}d\theta'^{2}+r'^{2}\sin^{2}\theta'd\phi'^{2}\right),
\end{array}\label{eq: ECEF frame 2}
\end{equation}
where \cite{key-5,key-20}
\begin{equation}
\Phi\equiv\frac{V}{c^{2}}-\frac{1}{2}\left(\frac{\omega_{E}r'\sin\theta'}{c}\right)^{2}.\label{eq: fi}
\end{equation}
represents the effective gravitational potential in the rotating frame.
If one applies the time scale change (\ref{eq: constant rate change})
in the non-rotating line element (\ref{eq: approximate solution})
and drops the primes on $t''$ in order to just use the symbol $t$,
one gets \cite{key-5,key-20}
\begin{equation}
ds^{2}=\left[1+\frac{2\left(V-\Phi_{0}\right)}{c^{2}}\right]\left(cdt\right)^{2}-\left(1-\frac{2V}{c^{2}}\right)\left(dr^{2}+r^{2}d\theta^{2}+r^{2}\sin^{2}\theta d\phi^{2}\right).\label{eq: approximate solution rescaled}
\end{equation}
The coordinate time $t$ in eq. (\ref{eq: approximate solution rescaled})
works in a very large coordinate patch which covers both the Earth
and the GPS satellite constellation \cite{key-5,key-20}. Hence, this
coordinate time can be used as a basis for synchronization in the
Earth's neighborhood \cite{key-5,key-20}. The difference $\left(V-\Phi_{0}\right)$
in the first term of Eq. (\ref{eq: approximate solution rescaled})
arises from the following issue \cite{key-5,key-20}. In the underlying
Earth-centered locally inertial frame where one uses the line element
(\ref{eq: approximate solution rescaled}), the unit of time is determined
by moving clocks in a spatially-dependent gravitational field \cite{key-5,key-20}.
One notices that in Eq. (\ref{eq: approximate solution rescaled})
both the effects of apparent slowing of moving clocks and frequency
shifts due to gravitation are present \cite{key-5,key-20}. As a consequence,
the proper time elapsing on the orbiting GPS clocks cannot be simply
used to transfer time from one transmission event to another \cite{key-5,key-20}.
Instead, path-dependent effects must be taken into due account \cite{key-5,key-20}
in perfect analogy with the discussion of clock synchronization in
Subsection 2.2.

We inserted the discussion in this Subsection in order to emphasize
that the obtained additional term $-\frac{1}{6}$ in Eq. (\ref{eq: z2})
is not an obscure mathematical or physical detail, but a fundamental
ingredient that must be taken into due account {[}4 - 6{]}.

\section{Erroneous criticism to our approach}

Our result {[}4 - 6{]} on the Mössbauer rotor experiment as new proof
of the GTR, that we have reanalysed in previous Section by correcting
an important mistake, has been criticized in \cite{key-17,key-18}.
We stress that such a criticism on our approach does not concern the
mistake that we corrected above. Instead, it concerns the issue that,
in the opinion of the authors of \cite{key-17,key-18}, the Mössbauer
rotor experiment cannot detect the second effect of clock synchronization.
The key point should be that the extra energy shift due to the clock
synchronization is of order $10^{-12}$... $10^{-13}$ and cannot
be detected by the detectors of $\gamma$-quanta which are completely
insensitive to such a very low order of energy shifts \cite{key-17}.
In addition, the authors of \cite{key-17} claim to have shown that
the extra energy shift can be explained in the framework of an alternative
gravitational theory proposed by themselves. They also insinuate that
such a new theory should replace the GTR as the correct theory of
gravity \cite{key-17,key-18}. Actually, we have recently show in
\cite{key-22} that the theory proposed in \cite{key-17,key-18} is
unscientific because, being a non metric theory, it macroscopically
violates EEP, which has today a strong, indisputable, empiric evidence
\cite{key-22,key-23}. In addition, in \cite{key-24} we have also
shown that, contrary to the claims in \cite{key-25}, the theory proposed
in \cite{key-17,key-18} can reproduce neither the LIGO\textquoteright s
\textquotedblleft GW150914 signal\textquotedblright{} nor the other
LIGO's detections of gravitational waves. 

Returning to the Mössbauer rotor experiment, in \cite{key-6} we have
shown that in \cite{key-17} the authors had a misunderstanding of
our theoretical analysis in \cite{key-5}. In fact, in {[}4 - 6{]}
and in Subsection 2.2 of the present paper, it has been shown that
electromagnetic radiation launched by the central source of the apparatus
is shifted of a quantity $0.\bar{6}\frac{v^{2}}{c^{2}}$ when arriving
to the detector of $\gamma$-quanta. Hence, the extra energy shift
due to the clock synchronization is of the same order of magnitude
of the first effect due to the ``gravitational blueshift'', as it
is obvious if one confronts Eq. (\ref{eq: gravitational redshift})
with Eq. (\ref{eq: z2}). In fact, in \cite{key-6} we have shown
that, with some clarification, our results in {[}4 - 6{]}, and, in
turn, in Subsection 2.2 of the present paper, hold also when we consider
the various steps of the concrete detection. In that case, the resonant
absorber detects the energy shift and the separated detector of $\gamma$-quanta
merely measures the resulting intensity, see \cite{key-6} for details.
In fact, as we clarified in \cite{key-6}, in {[}4 - 6{]} and in Subsection
2.2 of the present paper, we assumed as being negligible the difference
between the radial coordinate of the resonant absorber and the radial
coordinate of the detector of $\gamma$-quanta. In \cite{key-6} we
also clarified the meaning of Eq. (\ref{eq: z totale}). It represents
the total energy shift that is detected by the resonant absorber as
it is measured by an observed located in the detector of $\gamma$-quanta,
i.e. located where we have the final output of the measuring \cite{key-6}.
This is different from the total energy shift that is detected by
the resonant absorber as it is measured by an observed located in
the resonant absorber, which, instead, is given by Eq. (\ref{eq: gravitational redshift})
\cite{key-6}. The two quantities should be indeed equal \emph{only
}if the detector of $\gamma$-quanta should be rotating together with
the resonant absorber \cite{key-6}. Instead, the detector of $\gamma$-quanta
is fixed \cite{key-6}. The actual detector (i.e., the receiver of
electromagnetic radiation) is the resonant absorber, whose resonant
line is shifted with respect to the resonant line of the source in
the rotating frame. This induces the variation of intensity of resonant
$\gamma$-quanta, passing across this absorber \cite{key-2}. This
intensity is measured by the detector of $\gamma$-quanta, resting
outside the rotor system \cite{key-2}. The latter detector is rather
a technical instrument. It allows experimentalist to judge about the
shift of the lines of the source and the absorber via the measurement
of resonant absorption \cite{key-2}. But the key point is that the
shift of the lines of the source and the absorber that is observed
by an observer located in the rotating resonant absorber \emph{is
different} from the shift of the lines of the source and the absorber
that is observed by an observer located in the fixed detector of $\gamma$-quanta
\cite{key-6}. That difference is given by the additional factor $-\frac{1}{6}$
in Eq. (\ref{eq: z2}), which comes from clock synchronization. In
\cite{key-6} we also clarified that we are still measuring the total
energy shift by using the resonant absorber instead of by using the
detector of $\gamma$-quanta as it has been claimed in \cite{key-17}.
But the key point is that such a total energy shift measured by an
observer located in the fixed detector of $\gamma$-quanta is different
from the one measured by an observer located in the rotating resonant
absorber \cite{key-6}. Despite we clarified the above basic issues
in \cite{key-6}, the authors of \cite{key-17} wrote a subsequent
paper \cite{key-18} by claiming that our analysis in \cite{key-6}
was wrong and that the summation of the two components of relative
energy shift of our Eq. (\ref{eq: z totale}) is definitely inadmissible.
Notice that, by making such a claim, the authors of \cite{key-18}
are implicitly stating that:
\begin{itemize}
\item An apparatus realized to measure a time dilation can measure a particular
time dilation but it cannot measure a second time dilation which has
the same order of magnitude of the first time dilation. We indeed
stress again that the effect of Eq. (\ref{eq: gravitational redshift})
\emph{has the same order of magnitude} of the effect of Eq. (\ref{eq: z2}),
contrary to the claims in \cite{key-17}.
\item The result of a classical (i.e. non-quantum) experiment depends on
the way in which the experiment is realized. 
\end{itemize}
Of course, both of the two statements above are completely unscientific.
In their attempt to justify such unscientific statements, the authors
of \cite{key-18} claim that our above assertion that ``\emph{the
total energy shift measured by an observer located in the fixed detector
of $\gamma-$quanta is different from the one measured by an observer
located in the rotating resonant absorber}'', the we originally stated
in \cite{key-6}, is inconsistent from the relativistic viewpoint,
and leads to a causal contradiction. They claim indeed that the result
of a measurement in Mössbauer rotor experiments is the number of pulses
$N$ (i.e., the number of detected resonant $\gamma-$quanta passing
through a resonant absorber) during a fixed time interval $T$ at
the given tangential velocity $v$ for the absorber. Thus, if one
considers an observer co-rotating with the absorber (called the frame
$K'$ in \cite{key-18}) and a laboratory observer (called $K$ in
\cite{key-18}) attached to the detector placed outside, a strong
causal requirement is the equality 
\begin{equation}
N=N',\label{eq: causal requirement}
\end{equation}
being $N'$ the number of pulses in the frame $K'$. This means that
the indication of a counter of the pulses connected with the output
of the detector represents an absolute fact \cite{key-18}. In other
words, it must be the same for any observer (including the introduced
observers in the frames $K$ and $K'$) \cite{key-18}. Stating this,
the authors of \cite{key-18} claim that the average number of counts
which they define as \cite{key-18} 
\begin{equation}
\bar{N}=\intop_{0}^{T}I(t)dt,\label{eq: erroneous}
\end{equation}
which is obtained via the averaging of the indications of the counter
after multiple successive measurements in the frame $K,$ would be
equal to the average number of counts 
\begin{equation}
\bar{N}'=\intop_{0}^{T'}I'(t')dt',\label{eq: erroneous 2}
\end{equation}
as seen in the frame $K'$. In the notations of \cite{key-18} $I(t)$
and $I'(t')$ are the intensities of the resonant radiation passing
through the resonant absorber in the frames $K$ and $K'$ respectively,
while, in the opinion of the authors of \cite{key-18}, the remaining
designations should be obvious. After this, the authors of \cite{key-18}
claim that the intensities $I(t)$ and $I'(t')$ can differ from each
other due only to the difference of $dt$ and $dt'$, which, in their
opinion, should be the time dilation effect between the frames $K$
and $K'$. Then, the authors of \cite{key-18} end their analysis
by claiming that for the Mössbauer rotor experiment, the latter has
a typical difference of the order $10^{-12}$... $10^{-13}$ for either
rotating or resting observers. Thus, in the opinion of the authors
of \cite{key-18}, the admissible range of relative difference between
$I(t)$ and $I'(t')$ should entail the same order of magnitude and
should be totally negligible \cite{key-18}. Actually, there are a
lot of confusion, mistakes and nonsense in the analysis of \cite{key-18}
(that we just reviewed above) which concerns the supposed causal contradiction.
Now, we clarify the situation by showing that there is no causal contradiction
instead. There are indeed various points to be clarified:
\begin{enumerate}
\item By using the notations $dt$ and $dt'$ it seems that the authors
of \cite{key-18} used coordinate times rather than proper times in
computing the number of pulses in the two different frames. Instead,
one must use the proper time. Thus, one must correctly rewrite Eqs.
(\ref{eq: erroneous}) and (\ref{eq: erroneous 2}) as 
\begin{equation}
\bar{N}=\intop_{0}^{T_{R}}I(\tau)d\tau,\label{eq: correct}
\end{equation}
and 
\begin{equation}
\bar{N}'=\intop_{0}^{T_{L}}I'(\tau)d\tau,\label{eq: correct 2}
\end{equation}
respectively, where $T_{R}$ is the proper time which is measured
by the rotating observer and $T_{L}$ is the proper time which is
measured by the observer at rest in the laboratory. By using Eqs.
(\ref{eq: gravitational redshift}) and (\ref{eq: z2}) one easily
gets 
\begin{equation}
T_{R}\simeq\tau_{1}\left(1-\frac{1}{2}\frac{v^{2}}{c^{2}}\right)\label{eq: tr}
\end{equation}
and 
\begin{equation}
T_{L}\simeq\tau_{1}\left(1-\frac{1}{2}\frac{v^{2}}{c^{2}}-\frac{1}{6}\frac{v^{2}}{c^{2}}\right)=\tau_{1}\left(1-\frac{2}{3}\frac{v^{2}}{c^{2}}\right),\label{eq: tl}
\end{equation}
where we recall that it is $\tau_{1}\simeq\frac{r'_{1}}{c},$ being
$r'_{1}\simeq c\tau_{1}$ the radial distance between the source and
the detector, see Subsection 2.1 of this paper.
\item The time dilation effect between the frames $K$ and $K'$ is NOT
given by the difference of $dt$ and $dt'$. In fact, it \emph{cannot
be} an infinitesimal time because, as we stressed in Section 2 of
this paper, in a relativistic process of clock synchronization time
variations \emph{cannot be} merely computed by transferring time from
one transmission event to another. Instead, path-dependent effects
must be taken into due account, see also \cite{key-5,key-20}. Thus,
by using Eq. (\ref{eq: z2}), the correct value of the time dilation
effect between the frames $K$ and $K'$ is given by 
\begin{equation}
T_{L}-T_{R}\simeq-\frac{\tau_{1}}{6}\frac{v^{2}}{c^{2}}.\label{eq: time dilation effect}
\end{equation}
\item The time dilation effect between the frames $K$ and $K'$ is NOT
of the order $10^{-12}$... $10^{-13}$ (without units!) for either
rotating or resting observers as it has been claimed by the authors
of \cite{key-18}. In fact, it is the relative ratio 
\begin{equation}
\frac{T_{L}-T_{R}}{\tau_{1}}\simeq-\frac{1}{6}\frac{v^{2}}{c^{2}},\label{eq: relative ratio}
\end{equation}
which is of the order $10^{-12}$... $10^{-13}$. Instead the time
dilation effect between the frames $K$ and $K'$ is of the order
\[
\left|T_{L}-T_{R}\right|\simeq\left|\left(10^{-12}...10^{-13}\right)\right|\times\tau_{1}.
\]
\end{enumerate}
Now, let us discuss the physical implications of points 1. - 3. The
fundamental issue is that, contrary to the claims of the authors of
\cite{key-18}, the time dilation effect between the frames $K$ and
$K'$ is NOT negligible. In fact, the authors of \cite{key-18} claim
that their Mössbauer rotor apparatus detects a total time dilation
effect 
\begin{equation}
\triangle\tau\simeq-\frac{2\tau_{1}}{3}\frac{v^{2}}{c^{2}},\label{eq: time dilation YARK}
\end{equation}
which can be explained through their proper gravitational theory.
But this total time dilation effect is of the same order of the time
dilation effect between the frames $K$ and $K'$, as it is shown
by Eq. (\ref{eq: time dilation effect}). Thus, if the quantity of
Eq. (\ref{eq: time dilation effect}) is negligible, also the quantity
of Eq. (\ref{eq: time dilation YARK}) \emph{must be negligible},
and this merely implies that the Mössbauer rotor apparatus should
not work! On the other hand, the issue that the time dilation effect
between the frames $K$ and $K'$ is not negligible does not imply
the presence of causal contradiction. Instead, it merely means that
the two different observers in the two different frames $K$ and $K'$
measure the \emph{same} number of pulses in \emph{different }intervals
of proper time. It is exactly this issue which generates the additional
effect of clock synchronization. One can also reason as it follows.
Let us assume that the authors of \cite{key-18} are correct and that
the difference $T_{L}-T_{R}$ is negligible. Then, in the opinion
of the authors of \cite{key-18} one should find
\begin{equation}
T_{L}\simeq T_{R}\simeq\tau_{1}\left(1-\frac{2}{3}\frac{v^{2}}{c^{2}}\right),\label{eq: tl tr}
\end{equation}
which is the value predicted by the gravitational theory proposed
by the authors of \cite{key-17,key-18} without considering the additional
effect of clock synchronization. But, in that case, one can use the
same erroneous argument of the authors of \cite{key-18} concerning
the supposed causal contradiction. In fact, in absence of rotation,
the number of pulses measured by the observer in the laboratory is
given merely by 
\begin{equation}
\bar{N}_{0}=I_{0}\tau_{1},\label{eq: correct 3}
\end{equation}
where $I_{0}$ is the intensity of the resonant radiation passing
through the resonant absorber in absence of rotation. Instead, if
one actives the rotation, the number of pulses measured by the observer
in the laboratory is given by Eq. (\ref{eq: correct}) or by Eq. (\ref{eq: correct 2}),
which now coincide because we are assuming that the difference $T_{L}-T_{R}$
is negligible. But, following the ill logic of the authors of of \cite{key-18},
if the quantity 
\begin{equation}
T_{L}-\tau\simeq-\frac{2}{3}\frac{v^{2}\tau_{1}}{c^{2}},\label{eq: tl-tau}
\end{equation}
is not negligible, then we should have a causal contradiction because
we should obtain $\bar{N}_{0}\neq\bar{N}.$ Of course, this is pure
nonsense, because the observer in the laboratory measures the \emph{same}
number of pulses in \emph{different } intervals of proper time in
the two different cases of absence or presence of rotation. 

All the other criticisms in \cite{key-18} to our general relativistic
approach to the Mössbauer rotor experiment depends on the basic mistake
of the authors of \cite{key-18} concerning the supposed causal contradiction.
Thus, also such additional criticisms must be rejected.

We also observe the following. The gravitational theory proposed by
the authors of \cite{key-17,key-18} predicts a total value of $\frac{2}{3}$
for the coefficient $k$ in equation (\ref{eq: k}). But, in \cite{key-17,key-18}
the effect of of clock synchronization is \emph{not} taken into due
account. Thus, by considering also this additional effect, if one
uses Eq. (\ref{eq: z2}), one gets the correct value of 
\begin{equation}
k=\frac{2}{3}+\frac{1}{6}=\frac{5}{6}\label{eq: YARK out}
\end{equation}
for the gravitational theory proposed by the authors of \cite{key-17,key-18}.
This result is in contrast with the experimental results in \cite{key-2}.
Thus, contrary to the claims in \cite{key-17,key-18}, the theory
proposed in such works is completely ruled out by the Mössbauer rotor
experiment. We note that the authors of \cite{key-17,key-18} recently
published a new work with a further, clumsy attempt to show that our
results on the Mössbauer rotor experiment are wrong \cite{key-30}.
The present paper shows that they are the results in \cite{key-30}
which are wrong instead.

\section{Conclusion remarks}

Our results \cite{key-5,key-6} on the Mössbauer rotor experiment
representing a new, strong and independent, proof of Einstein's GTR
received an Honorable Mention at the Gravity Research Foundation 2018
Awards for Essays on Gravitation \cite{key-4}. In this paper, a mistake
which was present in our previous computations {[}4 -6{]} on this
important issue has been corrected by deriving a rigorous computation
of the additional effect of clock synchronization. In the final Section
of the paper we have also shown that some recent criticisms on our
general relativistic approach to the Mössbauer rotor experiment are
incorrect. Thus, the new insights of this paper ultimately confirm
our important result. 

For the sake of completeness, we stress that, on one hand, our result
is also confirmed by a recent result in \cite{key-26} on the Mössbauer
rotor experiment. On the other hand, another recent result in \textit{\emph{\cite{key-27}}}\textit{
}\textit{\emph{has shown that a general relativistic analysis similar
to the one proposed in this paper works also for the Sagnac effect.}}


\begin{thebibliography}{10}
\bibitem{key-1}R. L. Mössbauer, Zeitschrift für Physik A (in German)
\textbf{151}, 124 (1958).

\bibitem{key-2}A. L. Kholmetskii, T. Yarman, O.V. Missevitch and
B. I. Rogozev, Phys. Scr. \textbf{79}, 065007 (2009).

\bibitem{key-3}W. Pauli, \emph{Theory of Relativity, }Pergamon Press,
London (1958).

\bibitem{key-4}C. Corda, Int. Journ. Mod. Phys. D \textbf{27}, 1847016
(2018). 

\bibitem[5]{key-5}C. Corda, Ann. Phys. \textbf{355}, 360 (2015). 

\bibitem[6]{key-6}C. Corda, Ann. Phys. \textbf{368}, 258 (2016).

\bibitem[7]{key-7}W. Kündig, Phys. Rev. \textbf{129}, 2371 (1963).

\bibitem[8]{key-8}A. L. Kholmetskii, T. Yarman and O. V. Missevitch,
Phys. Scr. \textbf{77}, 035302 (2008).

\bibitem[9]{key-9}H. J. Hay et al, Phys. Rev. Lett. \textbf{4}, 165
(1960).

\bibitem[10]{key-10}H. J. Hay, in Proc. 2nd Conf. Mössbauer Effect,
ed A Schoen and D M T Compton (New York: Wiley) p 225 (1962).

\bibitem[11]{key-11}T. E. Granshaw and H. J. Hay, in Proc. Int. School
of Physics, \textquoteleft Enrico Fermi\textquoteright{} (New York:
Academic) p 220 (1963).

\bibitem[12]{key-12}D. C. Champeney and P. B. Moon, Proc. Phys. Soc.
\textbf{77}, 350 (1961).

\bibitem[13]{key-13}D. C. Champeney, G. R. Isaak and A. M. Khan,
Proc. Phys. Soc. \textbf{85}, 583 (1965).

\bibitem[14]{key-14}A. Kholmetskii, T. Yarman, O. Missevitch, Eur.
Phys. Jour. Plus, \textbf{128}, 42 (2013).

\bibitem[15]{key-15}V. O. de Haan, Jour. Comp. Meth. Scien. Eng.,
\textbf{13}, 51 (2013).

\bibitem[16]{key-16}A. L. Kholmetskii, T. Yarman, M. Arik and O.V.
Missevitch, AIP Conf. Proc. \textbf{1648}, 510011 (2015).

\bibitem[17]{key-17}T. Yarman, A. L. Kholmetskii, and M. Arik, Eur.
Phys. Jour. Plus \textbf{130}, 191 (2015).

\bibitem[18]{key-18}A. L. Kholmetskii, T. Yarman, O. Yarman, M. Arik,
Ann. Phys. \textbf{374}, 247 (2016).

\bibitem[19]{key-19}C. W. Misner , K. S. Thorne, J. A. Wheeler, \textquotedblleft Gravitation\textquotedblright ,
Feeman and Company (1973).

\bibitem[20]{key-20}N. Ashby, Liv. Rev. Rel. \textbf{6}, 1 (2003). 

\bibitem[21]{key-21}L. Landau  and E. Lifsits -\textit{ Classical
Theory of Fields} (3rd ed.). London: Pergamon. ISBN 0-08-016019-0.
Vol. 2 of the Course of Theoretical Physics (1971).

\bibitem[22]{key-22}C. Corda, Eur. Phys. J. Plus \textbf{133}, 456
(2018).

\bibitem[23]{key-23}C. M. Will, Living Rev. Rel. \textbf{17}, 4,
(2014). 

\bibitem[24]{key-24}C. Corda, Symmetry \textbf{10}, 558 (2018).

\bibitem[25]{key-25}T. Yarman, A. L. Kholmetskii, O. Yarman, C. B.
Marchal, M. Arik, Can. Journ. Phys., \textbf{95} \textbf{(10)}, 963
(2017). 

\bibitem[26]{key-26}G. Iovane, E. Benedetto, Ann. Phys. \textbf{403},
106 (2019). 

\bibitem[27]{key-27}E. Benedetto, F. Feleppa, I. Licata, H. Moradpour
and C. Corda, Eur. Phys. J. C. \textbf{79}, 187 (2019).

\bibitem[28]{key-28}A. Einstein, Ann. Phys. \textbf{49}, 769 (1916).

\bibitem[29]{key-29}H. Weyl, \emph{Raum, Zeit, Materie} (Verlag von
Julius Springer, Berlin, 1923), 5te Auflage, p.322; see also the 7th
edition republished by J.Ehlers in Spinger-Verlag in 1988.

\bibitem[30]{key-30}A. L. Kholmetskii, T. Yarman, O. Yarman and M.
Arik, Int. Journ. Mod. Phys. D (2019), 10.1142/S021827181950127X
\end{thebibliography}
\end{document}